\documentclass[10pt,aps,prb,twocolumn,tightenlines,letterpaper,superscriptaddress,nofootinbib,floatfix]{revtex4-2}
\usepackage{amssymb,latexsym}
\usepackage{amsmath,amsbsy,bbm}
\usepackage{epsfig,bm,color,bbold}
\usepackage{graphicx,url,hyperref}
\usepackage{xcolor}
\usepackage{cancel}
\usepackage{enumitem}

\usepackage{lineno}

\begin{document}

\def\a{{\alpha}}
\def\be{{\beta}}
\def\d{{\delta}}
\def\D{{\Delta}}
\def\P{{\Pi}}
\def\p{{\pi}}
\def\e{{\varepsilon}}
\def\ep{{\epsilon}}
\def\g{{\gamma}}
\def\k{{\kappa}}
\def\l{{\lambda}}
\def\L{{\Lambda}}
\def\m{{\mu}}
\def\n{{\nu}}
\def\o{{\omega}}
\def\O{{\Omega}}
\def\S{{\Sigma}}
\def\s{{\sigma}}
\def\t{{\tau}}
\def\x{{\xi}}
\def\X{{\Xi}}
\def\z{{\zeta}}

\def\ol#1{{\overline{#1}}}
\def\c#1{{\mathcal{#1}}}
\def\b#1{{\bm{#1}}}
\def\eqref#1{{(\ref{#1})}}

\def\wt#1{{\widetilde{#1}}}

\def\ed#1{{\textcolor{magenta}{#1}}}
\def\edd#1{{\textcolor{cyan}{#1}}}


%

\author{Prabal~Adhikari}
\email{prabal.adhikari.physics@proton.me}
 \affiliation{
Physics Department,
        Faculty of Natural Sciences and Mathematics,
        St.~Olaf College,
        Northfield, MN 55057, USA}
\affiliation{
Kavli Institute for Theoretical Physics, 
	University of California, 
	Santa Barbara, CA 93106, USA}
\author{Brian~C.~Tiburzi}
\email{btiburzi@ccny.cuny.edu}
\affiliation{
Department of Physics,
        The City College of New York,
        New York, NY 10031, USA}
\affiliation{
Graduate School and University Center,
        The City University of New York,
        New York, NY 10016, USA}
\author{Sona~Baghiyan}
\email{baghiy1@stolaf.edu}
 \affiliation{
        St.~Olaf College,
        Northfield, MN 55057, USA}

%

\title{
Approaching the Thermodynamic Limit of an Ideal Gas
}

\begin{abstract}
For a gas confined in a container, 
particle-wall interactions produce modifications to the partition function involving the average surface density of gas particles. 
While such correlations have a vanishing effect in the thermodynamic limit, 
examining them is beneficial for a sharper understanding of how the limit is attained. 
We contrast a classical and a quantum model of particle-wall correlations within the canonical ensemble.
\end{abstract}

\maketitle

\section{Introduction}
\label{sec:introduction}

Among the primary goals of statistical mechanics is the demonstration 
of the underlying statistical foundation for the laws of thermodynamics.
Extensive thermodynamic quantities emerge for idealized systems containing a large number of particles, 
and many characteristic properties of macroscopic systems only appear in the thermodynamic limit%
~\cite{extensivity,thermolim}, 
such as phases and critical points. For an ideal gas with $N$ particles considered in this work, the thermodynamic limit is specified by 
$N \gg 1$, 
but with the particle number density 
\begin{equation}
\rho = N / V
\label{eq:rho}
,\end{equation}
held fixed.
Discussion of this thermodynamic limit is minimal in most undergraduate courses.
\footnote{
For a few paradoxes that emerge without proper treatment of the thermodynamic limit, 
see 
Ref.~\cite{10.1119/1.1621028}.
The equivalence between ensembles is another aspect of the thermodynamic limit which is often omitted, 
but is lucidly demonstrated for the microcanonical and canoncical ensembles in 
Ref.~\cite{equivalence}.
}
However, as the thermodynamic limit can only be an approximation to nature, it is useful to consider the size of corrections when the system varies from this limit.

The thermodynamic limit is well established for classical ideal gases in containers with inert walls 
and is a staple topic in statistical mechanics courses. 
Without particle-particle interactions in the gas, 
however,
particles-wall interactions are required to achieve thermal equilibrium. 
These interactions ensure that information about the initial preparation of the gas is washed out.  
After thermal equilibrium is attained, particle-wall correlations have a negligible effect on the gas, 
which is a feature characteristic of the thermodynamic limit.
The leading correction to the thermodynamic limit of a classical ideal gas can be gleaned qualitatively as follows. 
At equilibrium, 
the average number of particles interacting with the walls is proportional to the area of the container.  
For a gas of 
$N$ 
particles contained in a volume 
$V$, 
the area is proportional to 
$V^{2/3} \propto N^{2/3}$. 
The energy of the gas has two contributions:
the kinetic energy of the gas particles 
and the particle-wall interaction energy. 
The former is additive, that is,
$\propto N$,
due to the absence of particle-particle interactions, 
while the latter depends on the average surface density
of particles near the walls, $\propto N^{2/3}$.
The ratio of these effects 
is proportional to
$N^{-1/3}$ and therefore
vanishes as 
$N \to \infty$, 
resulting in thermodynamic quantities that are extensive to a good approximation.\footnote{For a discussion of finite size corrections to the free energy, see, for instance, Ref.~\cite{hill}.} 
For a gas of 
$10^{23}$
particles, 
extensivity is realized at better than one part in ten million.

The focus of the present work is to make the above argument concrete, 
at least in the context of static potential models, 
and thereby to sharpen student understanding of the thermodynamic limit.%
\footnote{
Another illuminating way to explore the approach to the thermodynamic limit is through numerical experiments, 
for example,
see 
Refs.~\cite{10.1119/1.3216470,10.1119/1.4822174,10.1119/1.4867489}.
} 
Particle-wall correlations are straightforwardly accounted for through interactions within the canonical ensemble, 
and can thus readily be addressed in undergraduate and graduate statistical mechanics courses.
In Sec.~\ref{sec:general}, 
we begin with generalities concerning the classical ideal gas partition function in a cubic volume
and contributions from the surface density of gas particles. 
Simple models for particle-wall interactions are taken up in 
Sec.~\ref{sec:models}. 
First,  
we use a simple interaction in a fully classical description of the gas. 
The leading effect of the interaction on the distribution of energy within the canonical ensemble is investigated, 
and \ed{it} matches the scaling with 
$N$
obtained from the qualitative argument above. 
Next, 
we use a quantum mechanical description of the gas. 
Enforcing Dirichlet boundary conditions on the single-particle wavefunctions 
provides a rudimentary particle-wall interaction of very short range.
Nonetheless, 
the same qualitative features of the approach to the thermodynamic limit are observed.   
At the end, 
Sec.~\ref{sec:conclusion}, 
we provide several suggestions for further investigations. 
Some of these can be adapted for problem sets, 
while others are suited as options for end-of-term projects.

\section{Confining a Classical Ideal Gas}
\label{sec:general}

For a system of fixed volume and with a fixed number of particles, 
the canonical ensemble characterizes its thermal equilibrium with a reservoir of heat. 
When the system is a classical gas of 
$N$
non-interacting particles, 
its canonical partition function 
$\c Z$
has the form
\begin{equation}
\c Z 
= 
\frac{\,\,Z^N}{N!}
\label{eq:Z}
,\end{equation}
where 
$Z$
is the single-particle partition function, 
which can be written as a phase-space integral as follows: 
\begin{equation}
Z = \int \frac{d^3 p \, d^3 r}{(2 \p \hbar)^3} 
\, e^{ - \be \, H(\vec{p},\vec{r}) }
\label{eq:spectral}
,\end{equation} 
where 
$H$
is the classical Hamiltonian governing a particle of the gas interacting with a confining potential,
and the parameter 
$\be = (k_B T)^{-1}$
is proportional to the inverse temperature of the reservoir. 
Included in the partition function is the Maxwell-Boltzmann factor
$N!$, 
which accounts for the classical statistics of identical particles
and is required so that intensive thermodynamic quantities possess a finite thermodynamic limit.
Additionally included is the factor 
$(2 \p \hbar)^3$,
which can be thought of as an elementary volume of points in phase-space; it renders the partition function dimensionless. 
Quantum mechanics allows
$2 \p \hbar$
to be identified with Planck's constant,
but
this is irrelevant in classical statistical mechanics. 

Throughout, 
the gas is confined in a cubical container of volume 
$V = L^3$. 
For an ideal gas, 
the functional form of 
$Z$
depends on the potential, encoded through the classical Hamiltonian, that confines the particles within the container. 
For simplicity, 
we assume that the confining potential respects the cubic symmetry of the volume. 
Due to this assumption, 
the single-particle partition function is the product of 
one-dimensional partition functions as follows:
\begin{equation}
Z = \left( Z_1 \right)^3
\label{eq:Zd}
.\end{equation}
With the further assumption of momentum-independent particle-wall interactions, 
the one-dimensional partition function itself factorizes into the product of a momentum contribution 
$P_1$
and a coordinate contribution
$Q_1$,
\begin{equation}
Z_1 = P_1 \, Q_1\,.
\label{eq:PQ}
\end{equation}
The momentum contribution is defined to be
\begin{equation}
P_1 = \int_{-\infty}^{+\infty} \frac{dp}{2 \p \hbar} \, e^{ - \be \frac{p^2}{2m}} = \frac{1}{\l_T}
,\end{equation}
where 
$m$
is the mass of a gas particle; 
and, 
by virtue of the $\hbar$, the inverse of the thermal de Broglie wavelength, defined below, appears,
\begin{equation}
\l_T
=
\hbar \,
\sqrt{\frac{2 \p}{m \, k_B T}}
\,
.\end{equation}
The coordinate contribution takes the form 
\begin{equation}
Q_1
=
\int_{-\infty}^{+\infty} 
e^{ - \be \, V(x)}
\,
d x
\label{eq:QQ1}
,\end{equation} 
where 
$V(x)$
is the one-dimensional particle-wall potential. 
The factorization in Eq.~\eqref{eq:PQ}
expresses the statistical independence of the momentum and position of the particle, 
which is a feature of classical statistical mechanics --- in particular the assumption that the interaction with the wall is independent of the particle's momentum.

In the typical treatment of the gas, 
a particle is confined through the potential 
\begin{equation}
V_c(x) 
= 
\begin{cases}
0, & 0  < x < L\\
+\infty, & \text{otherwise}
\end{cases}
\quad
\label{eq:Vc}
,\end{equation}
for which the coordinate contribution
$Q_1$
is simply 
$L$. 
The infinite height of the potential is necessary to confine the particle; 
however, 
the infinitely abrupt rise at each wall is rather unphysical.
For a realistic potential,%
\footnote{
For a treatment of confinement parameterized by values for the derivatives of the potential at the walls,
see
Ref.~\cite{10.1119/1.3386255}, 
which provides additional insight related to the equipartition theorem.
} 
we define an excluded length as
\begin{equation}
\ell(\be)
=
\frac{1}{2} 
\int_{-\infty}^{+\infty} 
\Big[
e^{- \be \, V_c(x)}
-
e^{- \be \, V(x)} 
\Big]
dx
\label{eq:ell}
.\end{equation}
When the potential 
$V(x)$
is given by 
Eq.~\eqref{eq:Vc}, 
the excluded length vanishes by construction; 
whereas, 
for a confining potential of finite range, 
$\ell(\be)$
effectively represents the length over which particle-wall interactions modify the gas. 
Intuitively, 
the relative importance of this range should be temperature dependent; 
for example, 
at high temperature,
the range of the confining potential should become negligible. 
This behavior is incorporated in the definition of 
$\ell(\be)$.

\begin{figure}
\begin{flushleft}
\includegraphics[width=0.425\textwidth]{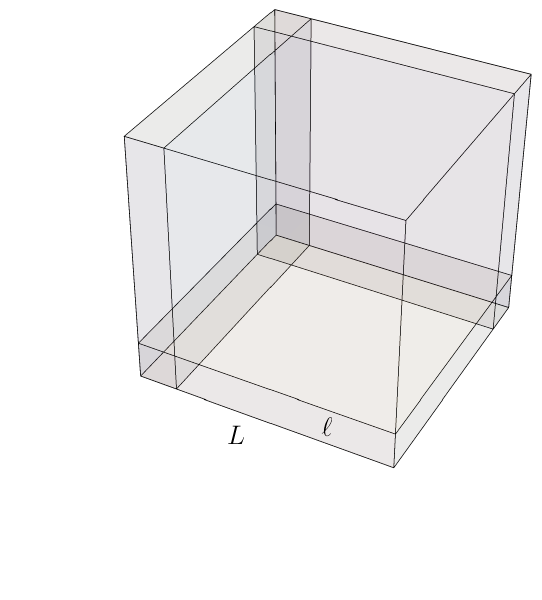}
$\phantom{space}$
\end{flushleft}
\vskip-6.5em
\caption{Cutaway of a cube of volume 
$L^3$
with a length 
$\ell$
depicted as excluded from each direction.
The three interior faces shown have volume 
$L^2 \times \ell$
excluded, 
except at the three edges shown where there is an overlapping volume of 
$L \times \ell^2$, 
aside from the far corner where there is an overlapping volume of
$\ell^3$. 
} 
\label{fig:cube}
\end{figure}

To see that the definition of
$\ell(\be)$ is a physically reasonable choice for describing the excluded length,
note that on account of Eqs.~\eqref{eq:PQ}--\eqref{eq:ell}
the one-dimensional partition function is given by
\begin{equation}
Z_1 = \frac{ L - 2 \, \ell(\be) }{\l_T} 
\label{eq:Ansatz}
.\end{equation}
This partition function is merely the ratio of the length accessible to a gas particle 
compared to 
$\l_T$, 
with 
$\ell(\be)$
excluded near each wall. 
Because the particle-wall interaction is independent of the size of the system, 
the excluded length must be held fixed in the thermodynamic limit. 
Since the box is cubical, the single-particle partition function is the cube of Eq.~\eqref{eq:Ansatz}
\begin{equation}
Z 
= 
\frac{L^3}{\l_T^3}
- 
6 \, \frac{L^2}{\l_T^2}  \frac{\ell(\be)}{\l_T}
+ 
12 \, \frac{L}{\l_T}  \frac{\ell^2(\be)}{\l_T^2}
- 
8 \, \frac{\ell^3(\be)}{\l_T^3} 
\label{eq:expand}
,\end{equation} 
where the terms are ordered according to their importance in the thermodynamic limit. 
Factors of 
$\left( L / \l_T \right)^d$
are
$d$-dimensional free-particle partition functions with zero-range confinement, 
and arise from integrals over the
$d$-dimensional phase space. 
Physical space is being excluded due to the finite range of confinement, 
while momentum space is unaltered. 
We can thus interpret each of the terms appearing in 
Eq.~\eqref{eq:expand}
with the aid of 
Fig.~\ref{fig:cube}
as follows.

\begin{enumerate}

\item
Volume:
the first term is the bulk contribution that survives the thermodynamic limit, 
which is the three-dimensional partition function with zero-range confinement. 

\item
Surface:
the second term accounts for the exclusion of the volume 
$L^2 \times \ell(\be)$
near each of the 
$6$ 
faces of the cube. 

\item
Edge:
the above exclusion over-subtracts contributions from the 
$12$ 
edges of the cube. 
Thus, 
the third term adds back the volume
$L \times \ell^2(\be)$
near each edge. 

\item
Corner: 
there is an over-addition of contributions from the 
$8$
corners of the cube, 
for which the last term subtracts the volume
$\ell^3(\be)$
near each corner. 

\end{enumerate}

We can rewrite the single-particle partition function
Eq.~\eqref{eq:expand}
in terms of the number density
$\rho$. 
Keeping only the leading correction to the thermodynamic limit, 
we have the expression
\begin{equation}
\frac{Z}{N} 
=
\frac{1}{\rho \,  \l_T^3}
\left[
1
-
6 
\left(
\frac{\rho}{N} 
\right)^{\frac{1}{3}}
\ell(\be)
+ 
\c O(N^{-\frac{2}{3}})
\right]
\label{eq:bulk+surf}
.\end{equation}
Consistent with dimensional analysis, 
a potential used to incorporate particle-wall interactions 
produces relative corrections 
$\propto N^{-1/3}$
that are due to the average surface density of gas particles.

From the canonical partition function, 
the distribution of energy within the gas can be investigated. 
Regardless of the size of
$N$, 
the statistical average energy
$\ol E$ 
can be computed from 
$\c Z$
using the formula%
\footnote{
At finite 
$N$, 
the average energy 
$\ol E$
is not identical to the thermodynamic internal energy 
$U$. 
The internal energy per particle 
$u$
is obtained by taking the thermodynamic limit of the average energy per particle
$u = \lim\limits_{N \to \infty}  \frac{\ol E }{N}$
at fixed 
$\rho$. 
Hence 
$\ol E$
differs from 
$U$
by finite-size effects. 
Similarly, 
the specific heat
$c_V = \big( \frac{\partial u}{\partial T} \big)_{N,V}$ 
is related to the variance in energy only in the thermodynamic limit, 
namely
$c_V =  -  \frac{\partial \be}{\partial T} \lim\limits_{N \to \infty} \frac{\s_E^2}{N}$.
For a general treatment of statistical fluctuations, 
see Ref.~\cite{callen}. 
} 
\begin{equation}
\ol E
=
-
\left(
\frac{\partial}{\partial \be}\log \c Z
\right)_{N,V}
,\end{equation}
because it is an energy-weighted sum of normalized probabilities.~\footnote{The $\log$ in the definition of the average energy is a natural $\log$.}
Using the general form of the partition function
Eq.~\eqref{eq:bulk+surf},
we readily obtain the expression
\begin{equation}
\ol E 
=
\frac{3}{2} N \, k_B T 
+ 
6 \, \rho^{\frac{1}{3}}  N^{\frac{2}{3}} \, \frac{\partial \ell(\be)}{\partial \be} 
+ 
\cdots
\label{eq:barE}
,\end{equation}
where only the leading correction to the thermodynamic limit has been retained. 
For a classical potential function, 
$\ell(\be)$
and hence 
$\ol E$
are necessarily independent of 
$\hbar$.
Statistical fluctuations in the energy of the gas are characterized by the variance in energy,%
~\cite{callen}
\begin{equation}
\sigma_E^2
\equiv
\overline{E^2} - \ol E {}^2
=
- \left(
\frac{\partial \ol E}{\partial \be} 
\right)_{N,V}
\label{eq:sigmaE}
,\end{equation}
and can be found from 
Eq.~\eqref{eq:barE}. 
Note that modifications to the distribution of energy away from the thermodynamic limit 
are sensitive to the temperature dependence of 
$\ell$. 
For this reason, 
we now turn to simple models for particle-wall interactions.

\section{Simple Models for Particle-Wall Interactions}   %
\label{sec:models}                                                          %

\subsection{Classical Potential Model}%

To model the confining potential, 
we introduce potential steps of height $V_{0}$
\begin{equation}
V(x) 
= 
\begin{cases}
V_0, & 0 < x < a \\
0, & a < x < L - a\\
V_0, & L- a < x < L \\
+\infty, & \text{otherwise}
\end{cases}
\quad
\label{eq:V0}
\end{equation}
as a simple way to approximate a confining potential of finite range.
The excluded length from each side of the container is calculated from 
Eq.~\eqref{eq:ell}, 
and leads to
\begin{equation}
\ell(\be)
=
a \left( 1 - e^{ - \be \, V_0} \right)
.\end{equation}
Notice that when 
$a = 0$, 
the potential reverts to 
Eq.~\eqref{eq:Vc}, 
and the excluded length accordingly vanishes. 
At high temperatures 
$k_B T \gg V_0$, 
the excluded length becomes 
$\ell = a \, \frac{V_0}{k_B T} + \cdots$, 
and vanishes as the temperature goes to infinity.

Using 
Eq.~\eqref{eq:V0} 
for the confining potential, 
the average energy in the canonical ensemble
computed from 
Eq.~\eqref{eq:barE},
\begin{equation}
\ol 
E
= 
\frac{3}{2} N k_B T
\left[
1
+
4 \, a \left( \frac{\rho}{N} \right)^{\frac{1}{3}} \frac{V_0}{k_B T} e^{- \be \, V_0}
+ 
\cdots
\right]
\label{eq:EbarClassical}
,\end{equation}
shows a small increase over the classical equipartition result. 
This increase is a direct effect of the repulsive step of height 
$V_0$
and width 
$a$. 
At high temperatures, 
the increase in average energy over equipartition is approximately constant
\begin{equation}
\Delta \ol E
=
6 N \, \frac{a}{L} \, V_0
,\end{equation}
and simply reflects the increase in energy from the potential 
$V_0$,
due to the average number of particles that experience it, 
i.e.,~those within a distance 
$a$
of the
six
cube faces.

\begin{figure}
\includegraphics[width=0.475\textwidth]{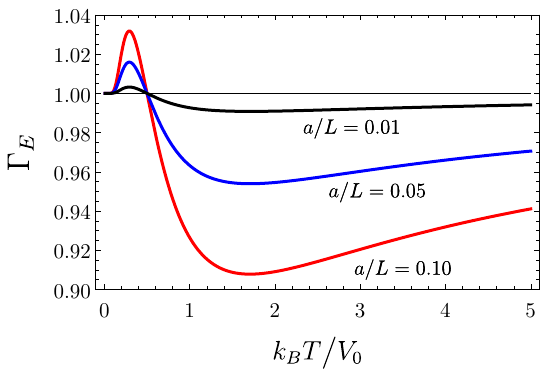}
\caption{Relative width of the energy distribution compared to a Gaussian
$\Gamma_E \equiv \frac{\s_E}{\ol E} \Big/ \left( \frac{\s_E}{\ol E} \right)_G$
as a function of temperature. 
Using a classical model of the confining potential
Eq.~\eqref{eq:V0}, 
the relative width 
Eq.~\eqref{eq:Gamma}
is shown
for three different values of the range parameter 
$a$,
spanning 
$1$--$10\%$
of the linear size of the container. 
As the temperature is raised past
$k_B T = \frac{1}{2} V_0$, 
the distribution changes from broader-than-Gaussian to narrower-than-Gaussian. 
} 
\label{fig:width}
\end{figure}

Using 
Eq.~\eqref{eq:sigmaE} 
to obtain the variance, 
the standard deviation in energy can then be compared to the average,
\begin{equation}
\frac{\s_E }{ \ol E}
=
\frac{
1
-
4 \, a \left( \frac{\rho}{N} \right)^{\frac{1}{3}} \frac{V_0}{k_B T} 
\left( 1 - \frac{V_0}{2 k_B T} \right)
e^{- \be \, V_0} + \cdots
}{\sqrt{\frac{3}{2} N \,} 
\label{eq:Gamma}
}
.\end{equation}
For a large number of particles, 
the energy distribution becomes sharply peaked about its average
$\ol E$. 
Interaction of the gas particles with the container walls modifies the 
$N^{-1/2}$
behavior of the relative width expected of a Gaussian distribution. 
Such particle-wall interactions
lead to correlations in the gas, 
for which there is a small deviation from the result for
$N$ 
completely independent distributions. 
There are two competing finite-size effects in the relative width of the distribution:
the average
$\ol E$
has increased \ed{and} the standard deviation
$\s_E$
has also increased. 
For 
$k_B T < \frac{1}{2} V_0$, 
the distribution is slightly broader than Gaussian 
(the standard deviation has increased more than the average); 
while, 
for  
$k_B T > \frac{1}{2} V_0$, 
the result becomes a little more sharply peaked than Gaussian  
(the average has increased more than the standard deviation). 
As the temperature is raised further, 
however,
any deviation from a Gaussian distribution becomes smaller, 
essentially washing out the particle-wall correlations. 
This behavior is depicted in 
Fig.~\ref{fig:width}, 
where the relative width is shown as a function of temperature.

\subsection{Quantum Mechanical Confinement}
\label{sec:quantum}

In the quantum mechanical treatment of the gas, 
in contrast to 
Eq.~\eqref{eq:spectral}, 
the single-particle partition function is the sum over energy eigenstates of the Boltzmann weights,
\begin{equation}
Z_1 
= 
\sum_{n = 1}^\infty e^{-  \be \, \e_n} 
\label{eq:Z1}
,\end{equation}
where
$\e_n$
is the 
$n$-th energy eigenvalue of the one-dimensionally confined particle. 
Enforcing 
Dirichlet boundary conditions,%
\footnote{
The effect of boundary conditions in the thermodynamic limit 
and the case of periodic boundary conditions are left for student investigations in 
Sec.~\ref{sec:conclusion}. 
} 
these energy eigenvalues are given by the standard particle-in-a-box formula as follows:
\begin{equation}
\e_n = \frac{\hbar^2}{2m} \, \left(\frac{ \p n}{L} \right)^2
.\end{equation}
The spectrum arises from the potential in 
Eq.~\eqref{eq:Vc}.
While this confinement has strictly zero range classically, 
quantum mechanical effects are non-local. 
A particle in the gas cannot be localized much better than its thermal de Broglie wavelength.  
We therefore expect that Dirichlet confinement quantum mechanically has a non-vanishing range on the order of 
$\l_T$.

Due to the dependence on 
$n^2$, the sum over positive values of $n$ in 
Eq.~\eqref{eq:Z1}
can be rewritten as a sum over positive and negative $n$
\begin{equation}
Z_1 = \frac{1}{2} \left(  \, \sum_{n = - \infty}^\infty e^{- \be \, \e_n} \right)  - \frac{1}{2}  
.\end{equation}
The term in parentheses is a sum, 
for which we have 
\begin{equation}
Z_1
= 
\frac{L}{\l_T}
\left[
1
-
\frac{\l_T}{2 L}
+ 
\c O \left( e^{- 4 \p \frac{L^2}{\l_T^2}} \right)
\right]
\label{eq:Zpower}
,\end{equation}
where the exponential corrections arise from replacing the sum with a corresponding Gaussian integral.%
\footnote{
While now seeming to have fallen out of favor, 
this result for the partition function had been standard in the study of statistical mechanics%
~\cite{fowler}. 
Accounting for the difference between sums and integrals using summation formulas has long been discussed%
~\cite{10.1119/1.1975156,10.1119/1.1976733,10.1119/1.1976672}, 
and has even been revisited more recently%
~\cite{Fernandez_2015,Ghosh_2015}.
}
These corrections are practically irrelevant. 
Even for a system with a very small size 
$L = 2 \, \l_T$, 
the exponential corrections amount to an effect on the order of   
$10^{-22}$. 
For such small systems, 
note that one would additionally need to account for the quantum statistics of gas particles. 
In what follows, 
the exponentially small corrections are ignored.

A simple interpretation%
\footnote{
This interpretation of the partition function is conceptually simple, 
but not mathematically unique.
For example, 
one could instead write the partition function
Eq.~\eqref{eq:Zpower}
in the form 
$Z_1 = \frac{L}{\L_T}$, 
where 
$\L_T = \frac{\l_T}{1 - \frac{\l_T}{2L}}$. 
Another simple interpretation could then be that the effective thermal de Broglie wavelength
$\L_T$
is larger 
$\L_T > \l_T$. 
This would imply that the average energy of a gas particle has decreased due to the imposition of 
Dirichlet boundary conditions.
Not only is it counterintuitive, 
this interpretation is proven false by 
Eq.~\eqref{eq:Ebar}. 
} 
of the result in Eq.~\eqref{eq:Zpower} is possible. 
Despite the quantum mechanical correlations between momentum and position, 
the partition function still factorizes
according to 
Eq.~\eqref{eq:PQ}. 
The coordinate contribution
$Q_1$, 
moreover, 
leads to the identification of the excluded length as
\begin{equation}
\ell(\be)
= 
\l_T / 4
,\end{equation}
which is indeed on the order of the thermal de Broglie wavelength. 
This purely quantum mechanical effect 
consequently
vanishes in the classical limit 
$\hbar \to 0$.

From the finite-length partition function given in 
Eq.~\eqref{eq:Zpower}, 
we obtain the average energy in the canonical ensemble as
\begin{equation}
\ol E
=
\frac{3}{2} N  k_B T
\left[
1 
+ 
\frac{\l_T}{2}
\left( \frac{\rho}{N} \right)^{\frac{1}{3}}
+
\cdots
\right]
\label{eq:Ebar}
,\end{equation}
which reflects a small increase in the average energy over the result of the classical equipartition theorem. 
Quantum mechanically, 
we expect an increase in this average due to the absence of zero-mode contributions  
in each of the three dimensions. 
The qualitative effect is the same as observed in the classical model
Eq.~\eqref{eq:EbarClassical}.
The physical origin, 
however,  
is different because there is no additional potential energy in the case of quantum mechanical confinement.  
The addition to the average energy appearing in
Eq.~\eqref{eq:Ebar} 
over the equipartition result can be written as
\begin{equation}
\D \ol E
=
6 N \, \frac{\l_T / 4}{L} \, \frac{k_B T}{2} 
,\end{equation}
which can be interpreted as an increase in the quadratic degrees of freedom, 
due to the average number of gas particles within 
$\ell(\be)$
of the 
$6$
cube faces.

Finally, 
we obtain the relative width of the distribution in energy by utilizing 
Eq.~\eqref{eq:sigmaE}
for the variance.
For quantum mechanical confinement
by Dirichlet boundary conditions, 
we find
\begin{equation}
\frac{\s_E }{ \ol E}
=
\frac{1 - \frac{3}{8} \l_T \left( \frac{\rho}{N} \right)^{\frac{1}{3}} + \cdots}{\sqrt{\frac{3}{2} N \,}}
\label{eq:Erelwidth}
.\end{equation} 
The distribution of energy is more narrowly peaked about the average compared to a
Gaussian distribution. 
Unlike the classical case, 
this is true for all temperatures; 
but, 
of course, 
any effect becomes washed out at high temperatures. 
In the quantum mechanical case, 
the particle-wall correlations disappear 
$\propto T^{-1/2}$, 
whereas in the classical case, 
the disappearance is 
$\propto T^{-1}$. 
The former behavior is due to the thermal de Broglie wavelength; 
whereas, 
the latter behavior is due to linear dependence on the model parameter
$V_0$, 
for which the correlations must then be proportional to
$\frac{V_0}{k_B T}$.

\section{Final Remarks and Further Directions}
\label{sec:conclusion}
In this work, we have investigated the approach to the thermodynamic limit of an ideal gas within the canonical ensemble. 
Using a classical potential model and quantum mechanical confinement through Dirichlet boundary conditions, we have characterized deviations of the average energy and energy fluctuations from the thermodynamic limit. 
Due to the absence of factorization of the partition function $\c Z$ in Eq.~\eqref{eq:Z} into a product of $N$ factors that are independent of $N$, the particles in the gas are correlated through $\rho$. 
A single particle does not know the gas density, which is an average property of many particles. 
Even when expressed in terms of $\rho$, the free energy per particle of an ideal gas maintains residual $N$ dependence due to particle-wall interactions.

While such dependence becomes negligible when $N\gg1$, there are various examples of small systems, not necessarily exemplifying particle-wall correlations, that are worth noting: 
whether the
$10^2$
nucleons of a nucleus can be used to define its temperature%
~\cite{Feshbach}, 
and the pioneering achievement of Bose-Einstein condensates
that first contained only
$10^2$
rubidium atoms
\cite{BEC1},
and soon after 
$10^4$
sodium atoms
\cite{BEC2}. 
Corrections to the thermodynamic limit are also relevant in numerical simulations of many-particle systems, 
which are necessarily carried out in a finite volume%
~\cite{tuckerman,PhysRevResearch.5.023156}. 
Additionally,
such corrections can be non-negligible for physical systems when the temperature is lowered, 
for which quantum mechanical finite-size effects are pronounced%
~\cite{10.1119/1.17416,10.1119/1.19048}.

To conclude, 
we provide several suggestions
for supplemental student investigations. 
These are organized based on their degree of difficulty, 
ranging from: 
questions, 
which can be addressed qualitatively; 
exercises, 
which are generalizations of the results obtained above;
projects, 
which require a more in-depth analysis;
to
challenges, 
which demand the most reasoning and adaptability. 

\subsection{Questions}%

\begin{enumerate}
\item 
In a model of confinement with finite range, 
what is problematic about holding 
$\ell / L$
fixed in the thermodynamic limit?

\item
Qualitatively describe the physical effects of choosing 
$V_0 < 0$
in the classical model for the confining potential 
Eq.~\eqref{eq:V0}. 
What might motivate such a choice?

\item
How do boundary conditions on the single-particle wavefunctions affect the thermodynamic limit?

\end{enumerate}

\subsection{Exercises}%

\begin{enumerate}

\item 
Model the confining walls with quadratic potentials, 
so that
Eq.~\eqref{eq:V0} 
is replaced by
\begin{equation}
V(x) 
= 
\begin{cases}
\frac{1}{2} m \o^2 x^2, & x < 0 \\
0, & 0 < x < L \\
\frac{1}{2} m \o^2 (x-L)^2, & x > L
\end{cases}
\,\,.
\notag
\end{equation}
Determine the excluded length 
$\ell(\be)$
classically, 
and assess the effect on the energy distribution.

\item 
A spherical container confines non-interacting gas particles with a potential of the form 
\begin{equation}
V(\vec{r} \,) 
= 
\begin{cases}
0, &  0 < r < R {-} a \\
V_0, & R {-} a < r < R \\
+ \infty, & r > R
\end{cases}
.\notag
\end{equation}
Analyze the approach to the thermodynamic limit of this gas using the
classical partition function.

\item 
Evaluate the quantum mechanical partition function 
Eq.~\eqref{eq:Z1} 
numerically and compare with the analytical formula given in 
Eq.~\eqref{eq:Zpower}. 
Carry out the analysis in terms of the dimensionless variable
\begin{equation}
\a = \frac{\p}{4} \left( \frac{\l_T}{L} \right)^2
.\notag
\end{equation}
In what regime of 
$\a$:
(\emph{i}) is the analytical formula valid,
and
(\emph{ii}) can quantum statistical effects be ignored?


\end{enumerate}

\subsection{Projects}   %

\begin{enumerate}

\item
Using summation formulas, 
analyze the quantum mechanical partition function 
Eq.~\eqref{eq:Z1}
when the non-interacting particles are subjected to periodic 
boundary conditions. 
How rapidly is the thermodynamic limit attained?

\item
Obtain the stated behavior of the quantum mechanical partition function 
Eq.~\eqref{eq:Zpower}
using 
the Poisson 
summation formula.

\item
Repeat the entire analysis for a two-dimensional ideal gas confined to an area
$L^2$
to address how the edges of the container modify the approach to the thermodynamic limit.

\end{enumerate}

\subsection{Challenges}%

\begin{enumerate}

\item
For a subsystem
$S$
that is energetically coupled to a reservoir 
$R$
of heat, 
what assumptions underlie the enumeration of accessible microstates according to the formula
\begin{equation}
\O (E_S, E) 
= 
\O_S(E_S) \, \O_R(E - E_S)
\notag
\, ?
\end{equation}
Define each of the quantities appearing above. 
How might this microcanonical description be modified to account for particle-wall interactions?

\item
Investigate the quantum mechanical partition function 
Eq.~\eqref{eq:Z1}
for the confining potential in 
Eq.~\eqref{eq:V0}. 
In approaching the thermodynamic limit, 
can both finite-range effects be obtained?

\end{enumerate}

\acknowledgments
%
P.A.~is supported in part by the U.S.~Department of Energy, the Office of Science, the Office of Nuclear Physics and the Quantum Horizons Program under Award No. DE-SC$0024385$.
P.A.~acknowledges the hospitality of the City College of New York, and the support of the Kavli Institute for Theoretical Physics, Santa Barbara, 
through which the research was supported in part by the National Science Foundation under Grant 
No.~NSF PHY-1748958. 
S.B.~acknowledges the support of internal funds through the Collaborative Undergraduate Research and Inquiry (CURI).
%

\section*{Author Declarations}
The authors have no conflicts to disclose.

\appendix
\bibliography{bibly}  %


\end{document}